\documentclass[preprint2]{aastex}
 \usepackage[dvips]{color}
\bibpunct{(}{)}{;}{a}{,}{,}
\usepackage{graphicx,float,amssymb,natbib,textcomp}
\shorttitle{NGC\,604 at NIR }
\shortauthors{Fari\~na et al.}

\begin{document}

\title{Unveiling the new generation of stars in NGC\,604 with Gemini-NIRI}
\author{Cecilia Fari\~na\altaffilmark{1} and Guillermo L. Bosch}
\affil{Facultad de Ciencias Astron\'omicas y Geof\'{\i}sicas, Universidad
Nacional de La Plata, 1900 La Plata, Argentina \\ IALP-CONICET, Argentina}
\and
\author{Rodolfo H. Barb\'a}
\affil{Instituto de Ciencias Astron\'omicas, de la Tierra y del Espacio
(ICATE-CONICET), Av. Espa\~na Sur 1512, (J5402DSP) San Juan, Argentina\\
Departamento de F\'{\i}sica, Universidad de La Serena, Cisternas 1200
Norte, La Serena, Chile}

\altaffiltext{1}{ceciliaf@fcaglp.unlp.edu.ar}

\begin{abstract}
We present a near infrared study focused on the detection and characterization of the youngest stellar component of the NGC\,604 giant star-forming region, in the Triangulum galaxy (M\,33). By means of color-color diagrams derived from the photometry of {\it JHK$_s$} images taken with Gemini-NIRI, we have found 68 candidate massive young stellar objects. 
The spatial distribution of these sources matches the areas where previous studies suggested that star formation might be taking place, and the high spatial resolution of our deep NIRI imaging allows to pinpoint the star-forming knots. An analysis of the fraction of objects that show infrared excess suggests that the star formation is still active, supporting the presence of a second generation of stars being born, although the evidence for or against sequential star formation does not seem to be conclusive. 

\end{abstract}


\section{INTRODUCTION}

In this paper we present a near infrared (NIR) photometric study of the NGC\,604
star-forming region. This study is focused on the detection and first
characterization of the NGC\,604 newest generation of stars. NGC\,604 is one of
the most outstanding giant H{\sc ii} regions (GHRs) in the Local Group. With a
L$_{H\alpha}$\,=\,2.6$\times$10$^{39}$ erg\,s$^{-1}$ \citep{2002MNRAS.329..481B} it is the
second most luminous H{\sc ii} region after 30 Doradus. The observation and study of
these two regions play a fundamental role in our understanding of the
astrophysical processes associated with massive stars and, in particular,
massive star formation and the environments where it takes
place. At a distance of 840 kpc, NGC\,604 is located far enough to observe global
characteristics but still close as to resolve individual objects and small
structures (1$\arcsec$\,$\sim$\,4 parsecs). NGC\,604 and 30 Dor provide the link between Galactic GHRs, which we are able to study in full detail, and distant GHRs, whose global properties can be traced to cosmological distances. Although they share several common
characteristics they also exhibit substantial differences such as their spatial structure and
ages distribution.

The ionizing stellar cluster is composed of a population of at least
200 O-type stars \citep{1996ApJ...456..174H}. Wolf-Rayet (WR) and Of stars, or
candidates, were first identified and classified in studies by
\cite{1981ApJ...249..471C}, \cite{1981ApJ...248.1015D}, \cite{1982A&A...108..339R}, 
\cite{1987MNRAS.226...19D}, and \cite{1998ApJ...505..793M}. New WR objects were
detected by \cite{1993AJ....105.1400D} by means of Hubble Space Telescope (HST) images, and an accurate
classification of known WR stars was made in \cite{2008MNRAS.389.1033D}.
\cite{1996MNRAS.279.1219T} spectroscopically classified two objects, one of them is a transition star from Luminous Blue Variable (LBV) to WR and the other is a red supergiant (RSG) star. There are forty OB stars with accurate spectral
classification by \cite{2003AJ....125.3082B}, three of them exhibit Of/WR
signatures (wind profile UV emission lines of N\,{\sc v}, C\,{\sc iv}, and Si\,{\sc iv} together with  strong emission line of He\,{\sc ii} at $\lambda$1640). \cite{2011MNRAS.411..235E} studied
the known evolved stellar population in the region by means of Spectral Energy Distribution (SED) fitting of HST {\it UVIJHK} photometry. The authors used the {\it UVI} photometric measurements from \cite{1996ApJ...456..174H} and they performed NIR photometry using archive images from HST-NICMOS (NIC2) in the F110W, F160W, and F205W filters. The latter HST data set, was originally used by \cite{2009Ap&SS.324..309B} to perform a photometric study of the region, in which they found five RSG candidates and a dozen candidate massive young stellar objects (MYSOs). 
However, the selection criteria for IR-excess sources was based on the maximum errors present in the NICMOS photometry. This suggests that a considerable number of sources may yet be detected with deeper images, using photometric errors appropriate for each object.

 The age of the main stellar population was estimated by several authors. \cite{1996ApJ...456..174H} analyzed the color-magnitude diagrams obtained from {\it UVI} HST-WFPC2 photometry; the authors fitted isochrones and considered the age constrains imposed by the presence of WR candidate stars. They concluded that the average ages for the stars in NGC\,604 range from 3 to 5 Myr, and that the presence of RSG could suggest the existence of an older subpopulation. This older subpopulation was later studied by \cite{2011MNRAS.411..235E} who concluded that the RSGs belong to a formation episode that occurred 12.4$\pm$2.1 Myrs ago. \cite{2000MNRAS.317...64G} analyzed stellar wind resonance lines in ultraviolet spectra taken with the International Ultraviolet Explorer (IUE), together with the nebular emission lines and the higher order terms of the Balmer series and He\,{\sc i} lines in absorption (both obtained from optical spectra taken with the 4.2-m William Herschel Telescope). The authors applied three different techniques of evolutionary synthesis and photoionization models (optimized for young star-forming regions) and concluded, consistently by the three methods, that the main ionizing cluster has an age of $\sim$\,3 Myr.
  The physical and kinematical properties of NGC\,604 were studied in \cite{2000PASP..112.1138M} (and references therein), by means of spectroscopic data obtained with the WHT (in particular, the lines formed in the warm ionized gas) and images from HST-WFPC2. The author concluded that NGC\,604 suffered one large starburst 3.0-3.5 Myr ago that formed the main ionizing cluster in the region.

 Regarding the stellar population spatial distribution in the region, it is worth mentioning that the objects in NGC\,604 are not centrally concentrated (as is the case for 30 Dor) but spread over a large projected area ($\sim$ 10000 pc$^2$ for
cluster A in \citealt{1996ApJ...456..174H}), composing a structure called the Scaled OB Association (SOBA) by \cite{2004AJ....128.1196M}.

According to the studies performed by \cite{2000ApJ...541..720T} and
\cite{2004AJ....128.1196M}, the interstellar medium exhibits two areas with
different excitation state: a high excitation region surrounded by a
low-excitation region, but not with a concentric geometry. These
structures also show different kinematic behavior and physical conditions. On
the west side (high ionization) there can be identified three cavities, which
seem to be bubbles bursting into the M\,33 halo (in our direction). 
There is a boundary, often cited as `the ridge', which extends along the north-south direction defining a clear east-west division within nebulae. The east side (low ionization) exhibits a quieter kinematic behavior. This dichotomy is also observed in the X-ray emission \citep{2008ApJ...685..919T}.

Detailed studies on the dynamics were also performed by
\cite{1996AJ....112..146Y}, who found that the width of the integrated velocity
profile needs the contributions of virial motion, thermal broadening, stellar
winds and SNRs. \cite{1997ApJ...487..163M} performed a study on the dynamics of
the region by means of the H$\alpha$ emission and they concluded that a few tens of WR
stars would be enough to feed the necessary kinetic energy. They also stated that the
turbulence is relatively young, giving another element in favor of recent processes of stellar formation taking place a few million years ago. Analyzing the velocity dispersion versus intensity diagram over the whole region, \cite{1996AJ....112.1636M} found that most velocity dispersion values measured in NGC\,604 are supersonic. The authors also used the same diagram as a tool to identify the zones corresponding to structures such as shells, loops or bubbles which are generated by the interaction of the massive stars' winds with the interstellar medium.

The NGC\,604 molecular component was widely studied by
\cite{1992A&A...265..437V}, \cite{1992ApJ...385..512W}, \cite{2003ApJS..149..343E}, \cite{2003ApJ...599..258R}, and \cite{2007ApJ...664L..27T}. In the latter study the
authors proposed a sequential star formation scenario triggered by the expansion
of the H{\sc ii} region. This idea was reinforced in a recent study with higher
spatial resolution in CO, HCN, and 89-GHz continuum emissions by
\cite{2010ApJ...724.1120M}.

In this study we present new observations of NGC\,604 acquired using the
Gemini-North telescope with NIRI in the {\it J}, {\it H}, and {\it K$_{short}$} ({\it K$_s$} hereafter)
bands. The high quality data obtained allowed us to perform a photometric study to identify and characterize candidate MYSOs in the ionizing cluster, confirming the candidates found by \cite{2009Ap&SS.324..309B} and increasing the sample presented in that study by a factor of five.
Our data and analysis methods are described in Section \ref{sec:data}. Results and implications are discussed 
in Section \ref{sec:results}.

\section{DATA REDUCTION AND ANALYSIS METHODS}
\label{sec:data}

\subsection{Observations}
The observations were made using the {\it Gemini-North} telescope with NIRI (Near
Infrared Imaging and Spectrometer) during the nights of September 08 and 11,
2005 (Proposal GN-2005B-Q-3). 
NIRI was used in the imaging mode with the f/6 camera. This configuration
yields a field of view of 120$^{\prime\prime}\times$120$^{\prime\prime}$ with a resulting
plate scale of 0.117 arcsec\,pixel$^{-1}$. The images were taken under
excellent seeing conditions, averaging 0.$^{\prime\prime}$35 FWHM measured in the {\it J},
{\it H}, and {\it K$_s$} images.
Table \ref{tab:filters} lists the broad-band
filters used, together with their central
wavelengths and wavelength coverage in Columns 2 and 3, respectively;
the exposure time for each individual image is given in Column 4, and in
Columns 5, 6, and 7 we have included details from the photometry that will be discussed in the following sections.

\begin{table*}
\centering
\caption{Main characteristics of the broad-band filters and the photometry}\label{tab:filters}\vspace{0.5cm}
 \begin{tabular}{ccccccccc} 
\tableline
Filter & $\lambda_c$ ($\mu$m) & Coverage ($\mu$m) & Exp. (s)  & Mean
mag\_err & Max. mag & Min. mag \\ 
 (1)   &     (2)              &  (3)              & (4)       & (5)   & (6)   & (7) \\ 
\tableline
{\it J} & 1.25 & 0.97-1.07 &  120 &   0.06 & 23.0      & 16.0  \\
{\it H} & 1.65 & 1.49-1.78 &  60  &   0.07 & 22.5      & 15.5  \\
{\it K$_s$} & 2.15 & 1.99-2.30& 40 &  0.10 & 22.0      & 15.5 \\ 
\tableline
\end{tabular}
\tablecomments{Column description: (1) broad-band filters used; (2) filter's central wavelength; (3) filter's wavelength coverage; (4) the exposure time for each individual image; (5) mean photometric uncertainties; (6) the photometric limiting magnitudes for an error smaller than 0.4 magnitude; (7) the magnitude at the saturation limit.}
\end{table*}

\begin{figure*}[]
\begin{center}
 \includegraphics[width=0.8\textwidth]{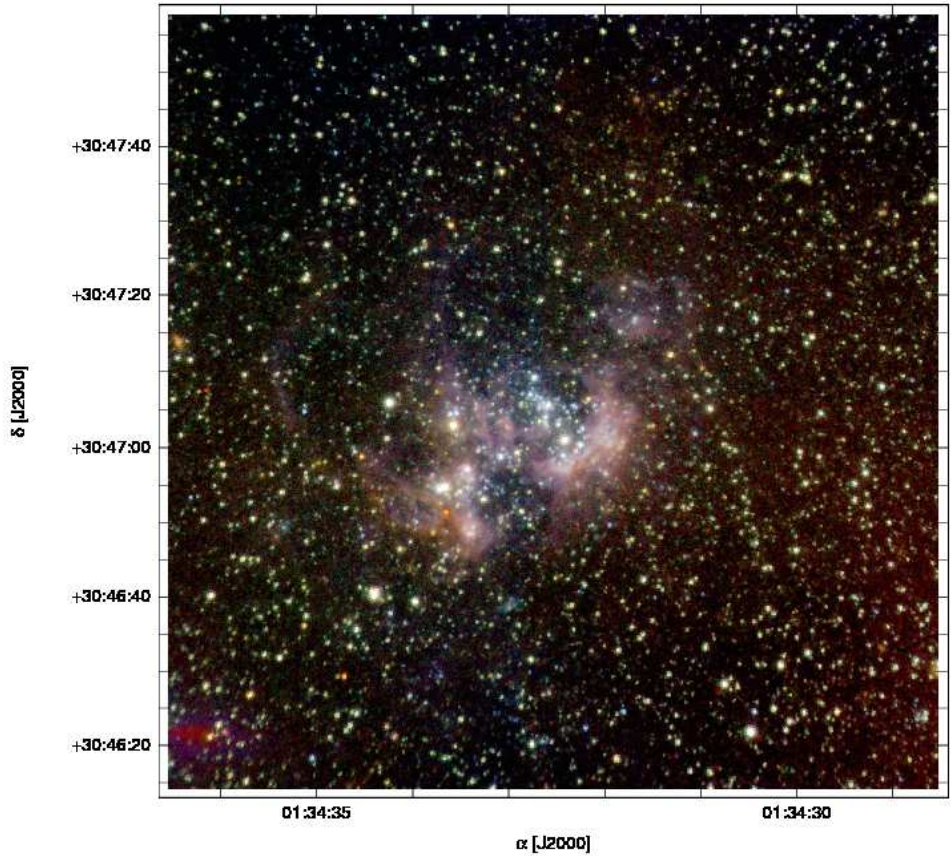}
 \caption{RGB composite image of NGC\,604 field: {\it J} (blue), {\it H} (green), and {\it K$_s$} (red). North is up, East to the left. Field size at NGC\,604's distance is roughly
490$\times$490 pc$^2$.}
   \label{fig:rgb}
\end{center}
\end{figure*}

 The observing sequences were made following
a square pattern with offsets of about 10$^{\prime\prime}$ every two consecutive field
images, interspersed with a sequence of sky images, also taken with dithering
offsets. The dithering sequence allows the removal of bad
pixels and other detector artifacts, although the effective area that gets fully exposed is smaller (107$^{\prime\prime}\times$107$^{\prime\prime}$). Figure \ref{fig:rgb} shows the color RGB image of NGC\,604 composed
from our final broad-band images: {\it J} (blue), {\it H} (green), and {\it K$_s$} (red).\\

\subsection{Data Reduction}

Most of the NIRI images exhibit a vertical striping pattern which originate
in the array's electronics. Although its spatial profile is well known (eight
columns wide), its intensity varies from one exposure to another, even within
an individual image, and it differs between the four image quadrants since each
quadrant in the NIRI array is read independently. The routine {\sc
nirinoise.py} (kindly provided by Gemini Staff)
was applied to each image (field images, sky images, flats on/off images, and
darks) in order to remove this pattern. Although it was not possible to
completely remove it in all individual images, the residual pattern, when it was present, was only of a few ADU and being a small spatial scale pattern, it was corrected in the final image, obtained by averaging the dithered images. 
The second correction was performed to account for the NIRI detector
non-linearity. This correction must be applied to all images with count rates
greater than $\sim$ 20 ADU s$^{-1}$. The routine,
also supplied by Gemini Staff,
 was applied to all {\it J}, {\it H}, and {\it K$_s$} field images.\\   
After these two initial image corrections the reduction procedure that follows was
pursued with {\sc IRAF}\footnote{{\sc IRAF} is distributed by the National Optical Astronomy Observatories,
    which are operated by the Association of Universities for Research
    in Astronomy, Inc., under cooperative agreement with the National
    Science Foundation.} routines in the {\sc gemini.niri} package. Following the instructions in the NIRI web pages, the images were sky-subtracted and
flattened. Short dark exposures were used to generate a mask to identify bad
pixels. Special care was taken when combining the individual images to obtain
the final image with the {\sc imcoadd} task. Due to possible confusion generated
by crowding and the highly variable background, the {\sc
imcoadd} align method was set in `user'  and the script
was called interactively twice: a first call to mark the reference objects in
each image and generate the transformation map, and a second call to generate all
the transformed images as well as to derive the average image, 
with bad pixels and flagged cosmic-ray hits omitted. The average image
was the final image used to perform the photometry.\\
Some images exhibited a spurious, large scale,
light pattern, which in most cases  increased the counts by no more than
0.5-1.0\% above the averaged background counts in a extended area of the field.
This effect seems to appear, every time, after the offset to observe the sky
region. Being a low spatial frequency pattern it is not expected to affect our
point-source photometry. Nevertheless, we examined its effect by generating two
final images for each filter, one including all the images and a second one including only  those images without any trace of the illumination pattern (reducing
the number of useful images by about 50\%). We performed photometric analysis
in both images and compared the results. As expected, the measured magnitudes
were the same for both images, confirming that they were not affected by the
light pattern, but the stacked image made with fewer individual images turned
out to be less deep and the measured magnitudes presented more scatter. We
therefore decided to pursue the study using the photometry derived from the
coadded image that included the images with the light pattern, discarding only a
few in which the pattern was more evident.     

\subsection{Photometry}

\subsubsection{PSF Fitting} 
\label{sec:PSF}
\begin{figure}
\begin{center}
  \includegraphics[width=0.49\textwidth]{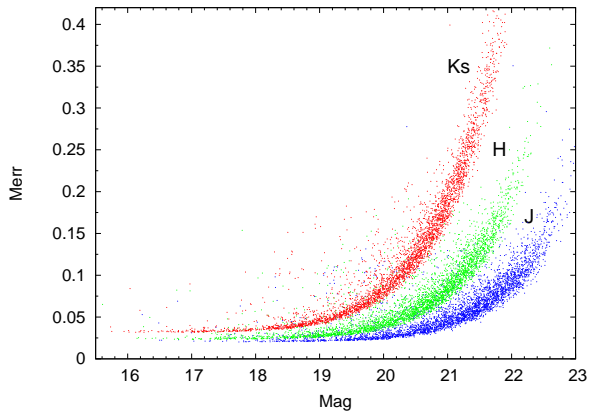}
 \caption{Error versus magnitude distribution for {\it J}, {\it H}, and {\it K$_s$} photometry. {\it J} band magnitudes show the smaller errors and {\it K$_s$} the relatively larger ones. This Figure is presented in a color version in the electronic edition of the Astronomical Journal.}
   \label{fig:mags}
\end{center}
\end{figure}

As evident in Figure \ref{fig:rgb}, the NGC\,604 field is highly crowded and the
background nebular emission exhibits strong variations on small spatial scales.
These conditions drove us to pursue Point Spread Function (PSF) photometry.\\
The objects for PSF fitting were selected using the {\sc daofind IRAF}
task from an image created by adding our {\it J} and {\it K$_s$} images. These images were aligned using the {\sc geomap} and {\sc geotran} IRAF tasks to derive and apply, respectively, the geometrical transformation (only linear shifts). By performing the object finding in the J+K$_s$ image we guaranteed the detection of stars in both extremes of the total wavelength range covered by the NIR
filters. A threshold of five sky sigmas over the background value
was set as the detection limit.\\ 
Stellar magnitudes were obtained by PSF fitting over the selected stars using the {\sc allstar} routine from {\sc DAOPHOT} software \citep{1987PASP...99..191S} in {\sc IRAF}. The
analytic PSF model that gave the best fit was a Moffat25 function (FWHM\,$\sim$\,3.0
pixels $\sim$\,0.$^{\prime\prime}$35) with parameters that could vary quadratically with position in the
image. Although a standard procedure, PSF construction and fitting in crowded
fields involves an iterative and careful process. As there are
no isolated stars available to be chosen as bona fide PSF stars to construct the
PSF model, it requires several sequences of PSF modeling and neighbor star
subtraction with an analytic PSF model of higher order each time, before the
final model can be obtained. Pixels used to determine the stellar profile were those with less
than 12000 ADU as this is the highest value for
which the linearity correction is expected to be reliable. The few stars that exceed this
limit in their central pixels had their magnitudes determined from
fits using the wings of their profiles.
The photometric uncertainties include the internal photometric error for each source as determined by DAOPHOT (considering  photon counts statistics, NIRI noise and gain characteristics and PSF-fitting errors), to which the aperture correction uncertainty and the error associated with the  transformation to the Mauna Kea Observatory (MKO) standard system were quadratically added. 
Aperture correction was calculated for each filter by performing aperture photometry (with the aperture used to measure the standard stars) over an average of $\sim$ 30 PSF stars with their neighbors previously subtracted. The error in the aperture correction is $\sim$ 0.02 mag. for the three filters. The uncertainties associated with the transformation of instrumental magnitudes to the MKO standard system are 0.08, 0.01, and 0.02 mag. for {\it J}, {\it H}, and  {\it K$_s$}, respectively.
In Table \ref{tab:filters} we have listed, for each filter, the mean photometric uncertainties in Column 5, the photometric limiting magnitudes for an error smaller than 0.4 mag in Column 6, and in Column 7 the magnitude at the saturation limit. Figure \ref{fig:mags} shows the error distribution with magnitude for each filter.\\

\subsubsection{Transformation to the Standard System}  

\begin{figure}[]
\begin{center}
  \includegraphics[width=0.49\textwidth]{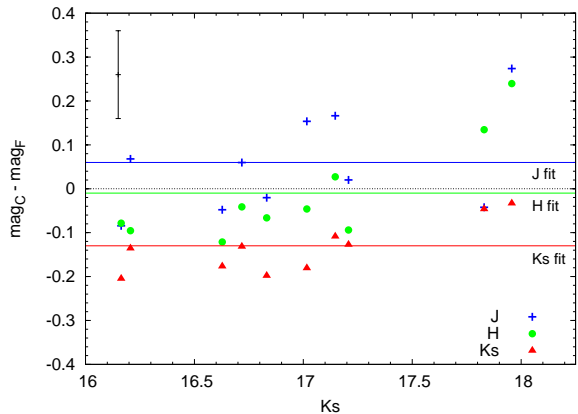}
 \caption{Comparison between our photometric results and \cite{2008A&A...487..131C} for the set of 10 stars in common. In the diagram there are plotted the magnitude differences for individual stars on each filter. Horizontal lines show the best fit for constant offsets at 0.06, -0.01, and -0.13 for {\it J}, {\it H} and {\it K$_s$}, respectively. The error bar in the top left corner represents the mean uncertainty for the magnitude differences. This Figure is presented in a color version in the electronic edition of the Astronomical Journal.} 
   \label{fig:cioni}
\end{center}
\end{figure}

The instrumental magnitudes were transformed to the MKO infrared photometric system, by means of aperture photometry performed on the standard stars. 
As the observed standard stars were limited to only one standard star per filter used in each observing night, we looked for a way to check the zero points of our magnitudes. A direct comparison with  our photometry and 2MASS photometry was not possible
as the magnitude limits of 2MASS at SNR\,=\,10 are 15.8, 15.1, and 14.3 for {\it J}, {\it H},
and {\it K$_s$} filters, respectively; which correspond to the magnitudes of stars that saturate in our images (see Table \ref{tab:filters}). 
Nevertheless, a comparison was possible by
means of a study published by \cite{2008A&A...487..131C}. The authors performed
wide-field {\it JHK$_s$} observations of M\,33 using WFCAM at UKIRT, referring the
astrometric and photometric calibration to the 2MASS point source catalog.
Fortunately, there are 10 sources in common with our photometry which allows us
 to compare their magnitudes with ours. The results of the comparisons are
shown in Figure \ref{fig:cioni} for the {\it J}, {\it H}, and {\it K$_s$} filters where the presence of a small magnitude offset between both sets can be seen. 
As no additional trend stood out as evident, and since the source number is small, we have limited our comparison to an evaluation of zero-point offsets. Performing a simple average to estimate these offsets we found them to be 0.06, -0.01, and -0.13 mag for {\it J}, {\it H}, and {\it K$_s$}, respectively. Although individually small, they add up to one tenth of a magnitude in the  ({\it H-K$_s$}) color. These offsets were added to the zero-points calculated for our photometry. The errors on these zero-point offsets to the magnitudes of the MKO system are smaller than the photometric uncertainties.\\

\subsubsection{Central and Field Regions. Photometric Uncertainties} \label{errors}

\begin{figure}
 \includegraphics[angle=-90, width=0.49\textwidth]{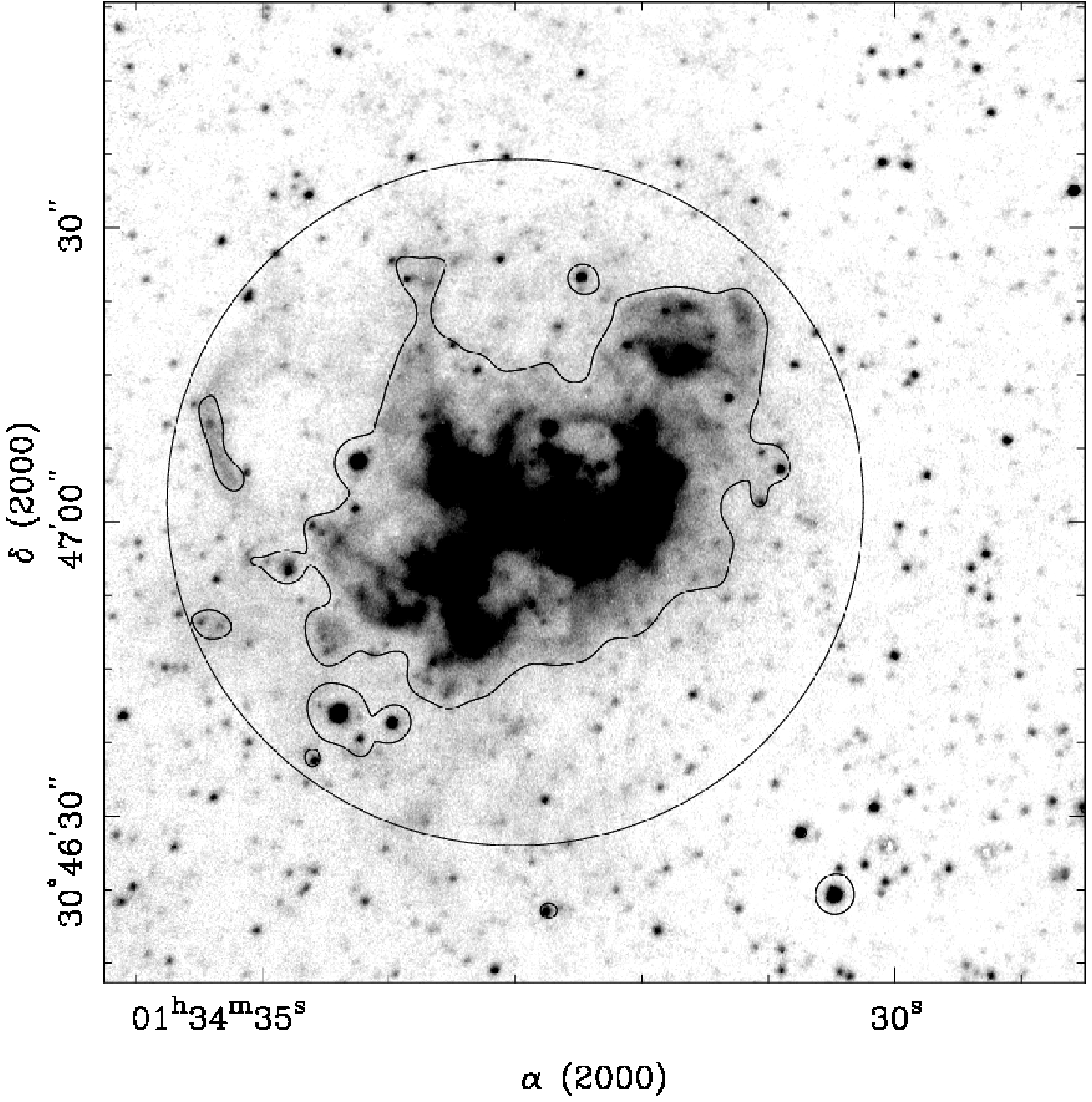}
  \caption{Grayscale Pa$\beta$ emission line image obtained with NIRI. The area of the central region in NGC\,604 is indicated with a circle that encompasses the contour at 5$\sigma$ level Pa$\beta$ emission, used to define the cluster radius. The area beyond the circle was used as the field region.}
   \label{fig:contornos}
 \end{figure}

\begin{figure}[]
\begin{center}
 \includegraphics[width=0.49\textwidth]{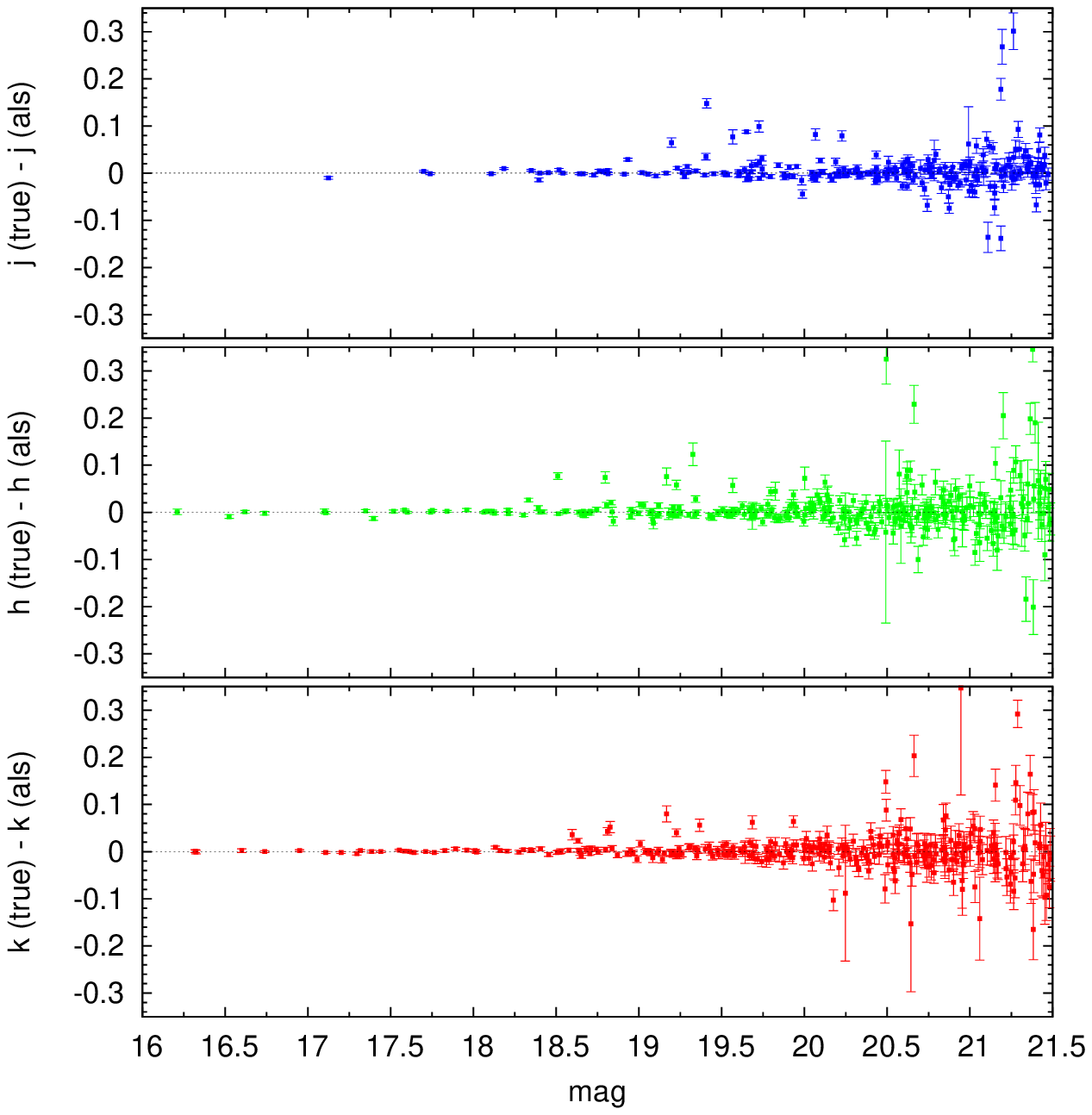}
 \caption{Results for {\it J} (top panel), {\it H} (middle panel), and {\it K$_s$} (bottom panel)
bands, from {\sc addstar} test to study the internal
photometric errors. The bars represent the magnitude uncertainties calculated by {\sc allstar} for
the measured magnitudes of the artificial stars. This Figure is presented in a color version in the electronic edition of the Astronomical Journal.}
   \label{fig:error_j}
\end{center}
\end{figure}

An analysis to examine the internal photometric uncertainties calculated by {\sc allstar}-DAOPHOT routine was made for each filter
by means of adding artificial stars to the science images. Our NGC\,604
field was divided in two regions: the `central region' and the `field region'. 
As the main cluster of NGC\,604 does not stand out as an evident increase in the stellar density, the central region limit was set
using a NIRI image obtained with the Pa$\beta$ narrow-band filter (details on narrow-band images and analysis will be included in a forthcoming paper). This is illustrated in
Figure \ref{fig:contornos} where the smoothed Pa$\beta$ contours at 5$\sigma$
used to define the central region limit are shown. The central region, enclosed by a $\sim$150 pc radius circle, is centered at $\alpha=01^h34^m33^s.14$ and $\delta=+30^{\circ}47\arcmin 1.\arcsec9$ and its area (of
68000 pc$^2$) encompasses the NGC\,604 SOBA. The region outside the circle, the field region, has a surface of $\sim$118000 pc$^2$ and was used to account for field star contamination.\\
A histogram of magnitude distribution (0.5 mag bin width)
was generated separately for each region. A new image was created by
adding artificial stars to each region. The number of added objects represented 10\%
of the stars at each magnitude interval. The artificial star
magnitudes were measured following the same procedure employed for the natural
stars. By comparing their measured magnitudes with their `true' magnitudes, we found that the differences are in the
range of the magnitude uncertainties calculated by the {\sc allstar} routine. In Figure
\ref{fig:error_j} we have plotted the
difference between the `true' and measured magnitudes for the artificial
stars for the {\it J}, {\it H}, and {\it K$_s$} filters in the top, middle, and bottom panels, respectively. The bars represent the magnitude error calculated by {\sc allstar} for
the measured magnitudes.

\subsubsection{Limiting Magnitudes}
\begin{figure}[!h]
 \includegraphics[width=0.49\textwidth]{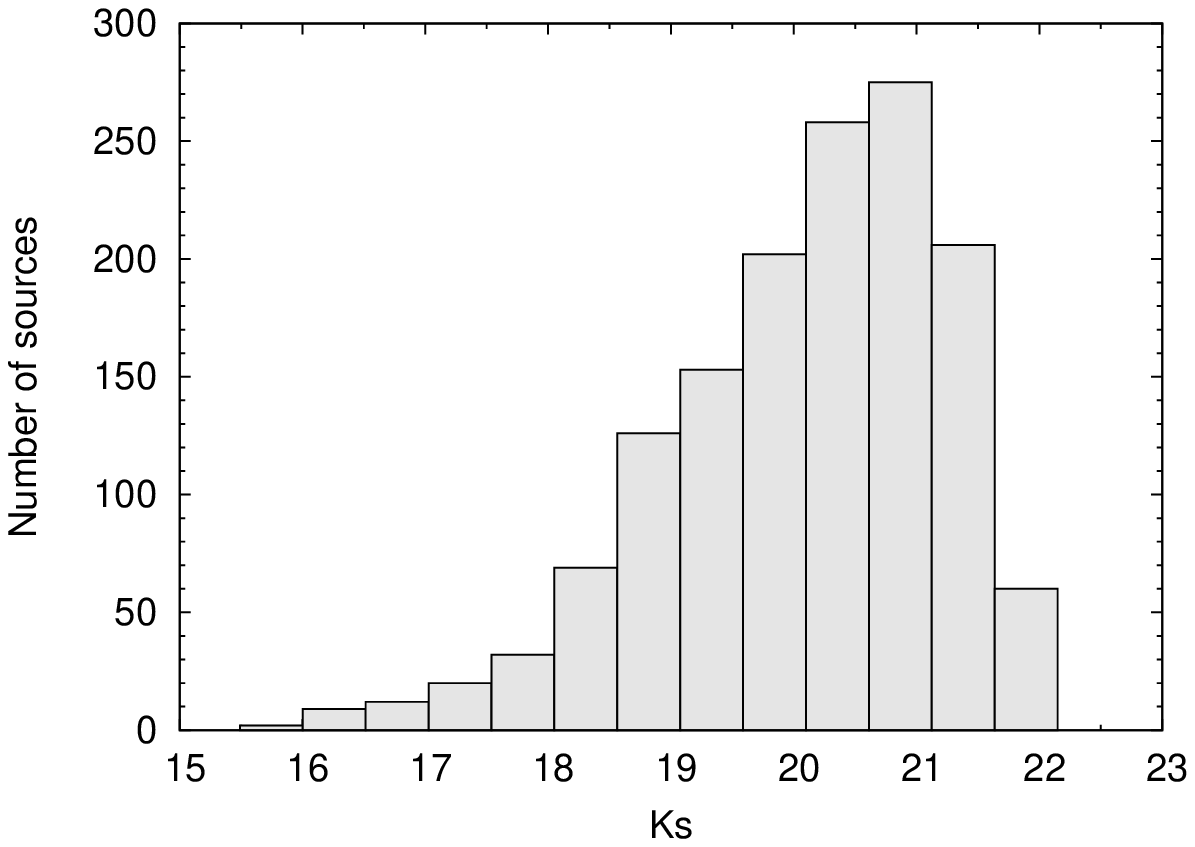}
 \caption{Histogram plot of the {\it K$_s$} magnitude distribution in the NGC\,604 central region. The limiting magnitude was derived from the histogram and it corresponds to the 20.5 mag bin, which comes right before the bin with the highest counts at 21.0 mag.}
   \label{fig:k_hist} 
\end{figure}

We have also used the magnitude histograms of the NGC\,604 central region to define the faintest magnitude reachable for each band. The magnitude limit was set equal to the magnitude of the first bin brighter than the peak of the histogram plot (e.g., see Figure~6). Following this criterion, the magnitude limits are 21.5 mag for {\it J} band and 20.5 mag for {\it H} and {\it K$_s$} bands.
Figure \ref{fig:k_hist} shows the histogram obtained for the {\it K$_s$} band where it can be observed that 20.5 mag bin comes right before the bin with the highest counts (21.0 mag). With these magnitude limits our sample is reduced to 1649 objects. However, the photometric uncertainties of the remaining objects are reduced to 0.04, 0.04, and 0.07 mag. for {\it J}, {\it H}, and {\it K$_s$}, respectively (nearly half of the mean magnitude uncertainties for the whole sample, see Table \ref{tab:filters}).

\subsection{Astrometry}
The astrometry was made in two steps, first an alignment in `image coordinates'
of the  final {\it J}, {\it H}, and {\it K$_s$} images; then a subsequent transformation from image coordinate system to World Coordinate System (WCS). The transformation from `image coordinates' to ($\alpha$,$\delta$) celestial equatorial
coordinates was performed using {\sc ccmap} and {\sc ccsetwcs} IRAF tasks, which computed the plate solution and created the WCS image, respectively. The astrometry was derived using 45 isolated stars, common to our images and version 2.3.2 of the Guide Star Catalog (GSC). Special care was taken to select well separated stars in our NIRI field. The reference frame of the GSC catalog is the International Celestial Reference Frame and Equinox J2000.0  \citep{2008AJ....136..735L}. The GCS 2.3.2 typical errors are 0.$^{\prime\prime}$3 and the transformation RMS are 0.$^{\prime\prime}$35 and 0.$^{\prime\prime}$25 for $\alpha$ and $\delta$,
respectively.

\section{RESULTS AND DISCUSSION}
\label{sec:results}

The magnitudes obtained in the individual filters were matched in a unique list
containing 3627 objects in the field in which all three {\it J}, {\it H}, and {\it K$_s$}
magnitudes were measured. Table \ref{tab:latabla} is an excerpt of the photometry table available in its entirety as machine-readable table in the electronic edition of the
Astronomical Journal. Column 1 contains the internal object id; Columns 2 and 3 the Celestial Equatorial coordinates (J2000.0) in degrees; Columns 4 and 5 the {\it J} magnitude and its uncertainty, respectively; Columns 6 and 7 the {\it H} magnitude and its uncertainty, respectively; Columns 8 and 9 the {\it K$_s$} magnitude and its uncertainty, respectively; and Columns 10 and 11 list the  ({\it J-H}) and  ({\it H-K$_s$}) colors, respectively. 
The total effective area - taking into account the dithering pattern - covered by our
photometry is of 107$^{\prime\prime}\times$107$^{\prime\prime}$ ($\sim$\,430$\times$430 pc$^2$ at the distance of M\,33).\\

\begin{table*}
\centering
\caption{Final photometry table}\vspace{0.5cm}\label{tab:latabla}
{\small
 \begin{tabular}{ccccccccccc} 
\tableline
Id& $\alpha$(J2000.0) & $\delta$(J2000.0) & {\it J} & $\sigma_J$ & {\it H} & $\sigma_H$ & {\it K$_s$} & $\sigma_{K_s}$&  ({\it J-H}) &  ({\it H-K$_s$})\\
   (1) &   (2)		& (3)		      & (4) & (5)      & (6) &   (7)    & (8)  &  (9)  &  (10) & (11)  \\
\tableline
2498 & 23.6167607 & 30.7957712 & 21.79 & 0.08 & 21.18 & 0.09 & 21.42 & 0.24 & 0.61 & -0.24 \\
4171 & 23.6168631 & 30.7972395 & 21.29 & 0.08 & 20.49 & 0.06 & 20.41 & 0.12 & 0.80 &  0.08 \\
3628 & 23.6169750 & 30.7976702 & 21.22 & 0.07 & 20.75 & 0.08 & 20.72 & 0.16 & 0.47 & 0.03 \\
2387 & 23.6170550 & 30.7946332 & 22.55 & 0.14 & 21.81 & 0.17 & 21.92 & 0.38 & 0.74 & -0.11 \\
2062 & 23.6171246 & 30.7913550 & 21.58 & 0.08 & 21.15 & 0.12 & 21.03 & 0.18 & 0.43 &  0.12 \\
\tableline
\end{tabular}
}
\tablecomments{Excerpt of the final photometry table available in its
entirety as machine-readable table in the electronic edition of the
Astronomical Journal. Column description: (1) internal object id; (2) and (3) Celestial Equatorial coordinates (J2000.0) in degrees; (4) {\it J} magnitude; (5) {\it J} magnitude uncertainty; (6) {\it H} magnitude; (7) {\it H} magnitude uncertainty; (8) {\it K$_s$} magnitude; (9) {\it K$_s$} magnitude uncertainty; (10) ({\it J-H}) color; and  (11) ({\it H-K$_s$}) color.}
\end{table*}

\subsection{Background/Foreground Stars}
\begin{figure}
 \includegraphics[width=0.49\textwidth]{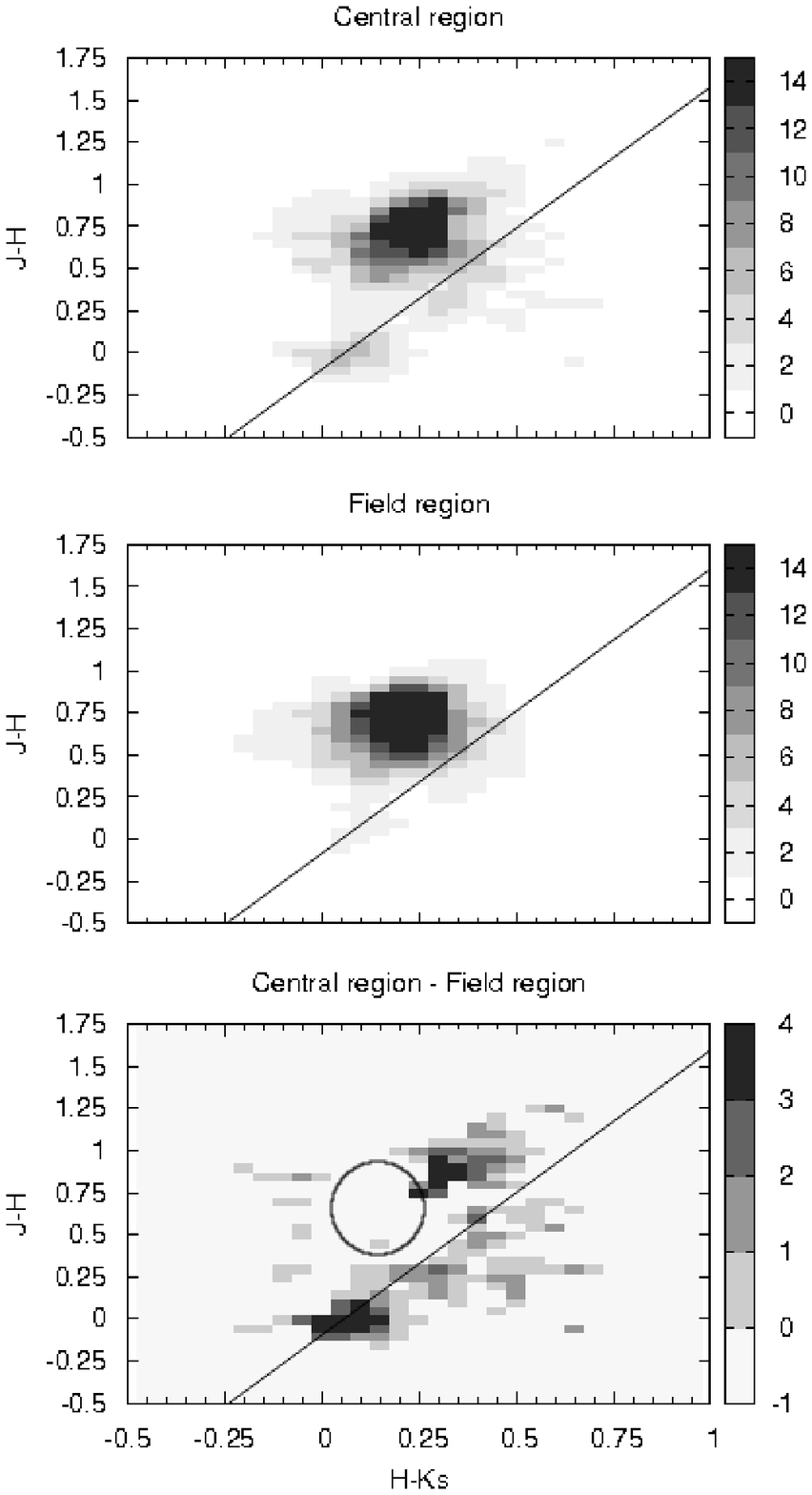}
 \caption{Density maps in the color-color diagram for objects in the central region of NGC\,604
(upper panel) and the control field (middle panel). The bottom panel shows the bin to bin subtraction resulting of the central region minus the field region. In this panel four features are evident, denoting the differences in the stellar populations of both regions: i) A relative over-density of early-type main-sequence objects around ({\it J-H})$\sim$0.0 and ({\it H-$K_s$})$\sim$0.0; ii) a relative over-density of red giant stars (that belong to M\,33's disk population) with moderate extinction around ({\it J-H})$\sim$1.0 and ({\it H-$K_s$})$\sim$0.25; iii) an area with negative counts, marked with a circle, due to the lack of unreddened red giant stars in the NGC\,604 region; iv) the IR-excess objects that lie to the right of the reddening line for a O6-O8 V star. These objects cover a larger area in the CC diagram of the NGC\,604 central region than for the field region, and the field shows almost no objects redder than ({\it H-$K_s$})$\sim$0.5.}
   \label{fig:control}
\end{figure}

As mentioned in Section 1, fortunately, NGC\,604 is situated in a privileged
location from an observational point of view: in a direction out of the Galactic
plane and in the outer part of M\,33 (a galaxy which is orientated almost face-on); hence,
it is expected that in the NGC\,604 field there is little contamination with
foreground and background stars. We have estimated the contamination by Galactic objects in our field using the model of \cite{2003A&A...409..523R}\footnote{By the electronic interface at http://model.obs-besancon.fr/}, which yields only six objects, most of
them F dwarf stars belonging to the thick disk.
Our main concern for the study presented in
this paper is the contamination by field objects lie in the area of the
color-color (CC) and color-magnitude (CM) diagrams occupied by MYSOs
candidates, resulting in possible misidentifications and an overestimation of
their number. To examine this effect we generated density plots (two
dimensional histograms) for the CC diagram of the NGC\,604 central region and the
field region surrounding it, as defined in the previous Section. 
Figure \ref{fig:control} exhibits the CC density maps for the central region
(top panel), the field region scaled to match the central region area (middle panel), and the bin to bin subtraction resulting of the central region minus the field region (bottom panel). Each bin in these diagrams is a 0.05$\times$0.05 mag. square, in ({\it J-H}) and ({\it H-$K_s$}). There are three visible differences between the object distribution in the CC
density maps from both regions, which become even more evident in the bottom panel: (a)
the star population around ({\it J-H})$\sim$0.0 and ({\it H-$K_s$})$\sim$0.0, which is present in the diagram of the central region but missing in the field region. These objects are located in the CC area belonging to early-type main-sequence (around O6-B2 V) and early-type giants
stars, hence they are expected to be part of the NGC\,604 population and do not suffer
noticeable reddening; (b) the number of objects that lie to the right of the reddening line for a O6-O8 V star, they constitute the IR-excess objects and are the main interest
for this study. It is clear that the IR-excess objects cover a larger area in the CC diagram of
the NGC 604 central region than for the field region, and the field shows almost no objects
redder than ({\it H-$K_s$})$\sim$0.5; (c) the location of the red giant stars in the observed CC diagram, grouped around ({\it J-H})$\sim$1.0 and ({\it H-$K_s$})$\sim$0.25, changes due to the reddening introduced by NGC\,604. This reflects in a relative over-density of reddened RGs when looking through NGC\,604 and a relative absence of their unreddened counterparts, seen as negative counts in the circled region marked in the bottom panel.\\

\subsection{Objects with IR Excess: IR excess Fractions and Spatial Distribution}

\begin{figure*}
\centering
 \includegraphics[width=0.5\textwidth]{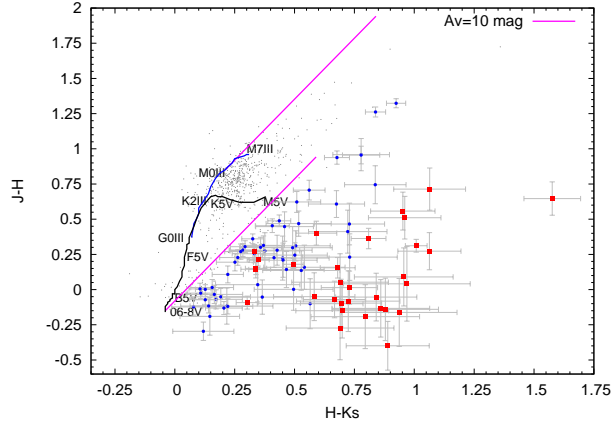}
 \caption{Color-color diagram for stars in the  central region of NGC\,604. Filled circles and squares denote objects with IR excess. Particularly, square symbols are objects that exhibit `extreme IR excess'. As a visual aid, the intrinsic colors of unreddened stars according to their spectral type and luminosity classes taken from \cite{2000asqu.book..143T} are also included. Straight lines indicate the reddening paths for the bluest and reddest stars, calculated considering an absorption of A$_V$=10.0 mag and the extinction coefficients by \cite{1985ApJ...288..618R}. Stars that lie to the right of the lower reddening line are those that show an infrared excess larger than one time and three times their photometric uncertainties, blue points and red squares, respectively (in the electronic color version of the paper). The error bars at $\pm$1 sigma in the ({\it J-H}) and ({\it H-K$_s$}) colors were also plotted for each object. This Figure is presented in a color version in the electronic edition of the Astronomical Journal.}
   \label{fig:CCD}
\end{figure*}

\begin{figure*}
\centering
 \includegraphics[width=0.5\textwidth]{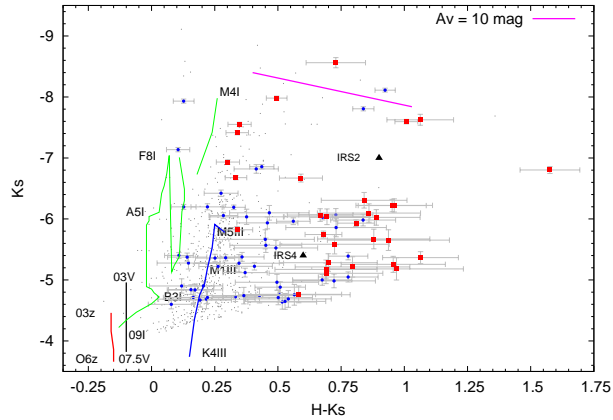}
 \caption{Color-magnitude diagram for stars in the  central region of NGC\,604. The same symbol codes as in Fig. \ref{fig:CCD} were used to flag stars that show IR excess. As a reference, we have added the expected location for stars according to their spectral types and luminosity classes, their ({\it H-K$_s$}) colors were taken the from \cite{2000asqu.book..143T}, and we have used the magnitudes of Main-Sequence stars from \cite{2006A&A...457..637M}, the magnitudes of Zero Age Main-Sequence stars from  \cite{1997ApJ...489..698H} and the magnitudes for supergiant stars from \cite{2007MNRAS.374.1549W}. Straight line indicates the reddening path calculated considering an absorption of A$_V$=10.0 mag and the extinction coefficients by \cite{1985ApJ...288..618R}. We have also included two well known MYSOs, namely S106\,IRS24 and NGC\,2024\,IRS4 (see text for discussion). The error bars at $\pm$1 sigma in the ({\it H-K$_s$}) color and the {\it K$_s$} magnitude were also plotted for each object. The distance modulus used (m$_V$-M$_V$\,=\,24.62) mag corresponds to M\,33's distance \citep{2001ApJ...553...47F}. This Figure is presented in a color version in the electronic edition of the Astronomical Journal.}
   \label{fig:CMD}
\end{figure*}

IR excess in MYSOs arise from heated dust and gas located in the very young objects'
surroundings, being a circumstellar phenomena rather than interstellar. This
material can be part of the original cloud material that was either left behind
after the protostellar collapse and not blown yet by the new born star or, it
can be material which forms part of the accretion disk of the forming object
still in the contraction process  \citep{palla}. Therefore, the objects' location in the CC diagram allows us to detect protostellar object  candidates, as sources that lie on the right side of the reddening line for a O6$-$O8 V star, exhibiting an IR excess. 
Figures \ref{fig:CCD} and \ref{fig:CMD} show the CC and CM diagrams obtained with our
photometry for the NGC\,604 central region. Those diagrams include 693 selected objects within the upper magnitude limits at {\it J}\,=\,21.5, {\it H}\,=\,20.5, and {\it K$_s$}\,=\,20.5 mag. 
As there is an intrinsic (and variable) uncertainty in the 
determination of NIR colors for each object, we have considered as `objects with IR excess' those sources whose distance to the reddening line is larger than their uncertainties (at one sigma) in the ({\it H-K$_s$}) color, marked with filled circles and squares in Figures \ref{fig:CCD} and \ref{fig:CMD}. In particular, the square symbols in both plots denote objects that exhibit `extreme IR excess', that is, their IR excess is larger than their errors by a factor of at least three.
In the CM diagram we have also included, as a reference, the location of two well known
Galactic objects: S106 IRS4, a Class 0 protostar \citep{1999ApJ...525..821F} and NGC\,2024 IRS2, originally 
identified as an Ultra-Compact H{\sc ii} region. Although \cite{2003A&A...404..249B} suggest that it might be a young massive star with a dusty disk, this type of object is still linked to very recent massive star formation.
With the criterion previously mentioned, we have selected a total of 88 objects that exhibit IR excess, 32 of them showing extreme IR excess. In Table \ref{tab:latablaIR} there are listed the 88 objects with IR excess, where Column 1 is the internal object id (as in Table \ref{tab:latabla}), Columns 2 and 3 the Celestial Equatorial coordinates (J2000.0) in degrees, and Columns 10 and 11 list the ({\it J-H}) and  ({\it H-K$_s$}) colors, respectively. This table is available in its entirety as machine-readable table,  in the electronic edition of the Astronomical Journal. \\

\begin{table}
\caption{Objects which exhibit IR excess.}\vspace{0.5cm}\label{tab:latablaIR}
{\small
 \begin{tabular}{ccccc} 
\tableline
Id& $\alpha$(J2000.0) & $\delta$(J2000.0) &   ({\it J-H}) &  ({\it H-K$_s$})\\
   (1) &   (2)		& (3)		      & (4) & (5)        \\
\tableline
4234 & 23.6323556 & 30.7843609 & 0.41 & 0.72 \\
4063 & 23.6325544 & 30.7835649 & 0.45 & 0.46 \\
3302 & 23.6326740 & 30.7833725 & 0.51 & 0.96 \\
4361 & 23.6328734 & 30.7839901 & -0.05 & 0.37 \\
4048 & 23.6329417 & 30.7830922 & 0.47 & 0.52 \\
\tableline
\end{tabular}
}
\tablecomments{\twocolumn{Excerpt of the table listing objects which exhibit IR excess. The table is available in its
entirety as machine-readable table in the electronic edition of the
Astronomical Journal. Column description: (1) internal object id; (2) and (3) Celestial Equatorial coordinates (J2000.0) in degrees units; (4) ({\it J-H}) color; and  (5) ({\it H-K$_s$}) color.}}
\end{table}

Besides MYSOs, in GHRs we can also find massive evolved stars like WR stars, B[e] supergiants (sgB[e]) and Of stars. As was commented in Section 1, the evolved stellar population
within NGC\,604 has already been widely studied. Particularly, the WR content of NGC\,604 was studied by \cite{1993AJ....105.1400D} with the HST/WFPC. In a complete survey, the authors identified 14 Of/WR candidates (up to M$_B$=22.88). From these 14 Of/WR candidates we detected 11 objects, since in our NIRI images objects WR4a - WR4b and  WR2a - WR2b could not be resolved, and WR11 is under the detection limit (indeed it is also optically faint). From the 11 WR detected in our photometry, 7 show IR excess greater than three times their uncertainty in ({\it H-K$_s$}) color, whereas the remaining four do not exhibit IR excess at all.
Therefore, from the 88 objects which exhibit IR excess, we have subtracted the 7 known WR stars. 

Regarding possible contamination with evolved B[e] stars, it is worth reviewing that the known objects with the B[e] phenomena are not abundant and most of the B[e] stars are relatively faint \citep{2006ASPC..355...13M}. From the five classes of B[e] stars \citep{1998A&A...340..117L}, we only expect to detect the most luminous objects (sgB[e]) at the distance of M\,33. Considering that the list of all known Galactic sgB[e] candidates in the study by \cite{2009A&A...494..253K} includes only 14 objects, we can infer that it is very unlikely to find a significant number of sgB[e] in NGC\,604. 

Summarizing, the NGC\,604 central region contains 693 objects with 81 showing IR excess (with the known  WR subtracted). There are 566 objects in the field region (after the scaling correction applied to match both surveyed areas), with 13 showing IR excess. Therefore, we can estimate a total number of 127 objects belonging to NGC\,604 in our photometry. From the 81 objects with IR excess in the region centered at NGC\,604, we subtracted 13 objects to account the background/foreground contamination, which yields 68 objects with IR excess in the NGC\,604 central region, and a fraction of objects with {\it JHK} IR excess, $\eta_{IR}$, of 68/127 ($\sim$\,54\%). 

As was previously mentioned in the Introduction, \cite{2009Ap&SS.324..309B} found a dozen candidate MYSOs by means of HST-NICMOS NIR data. In the present study we were able to detect ten of these objects and increase the sample of MYSOs candidates by a factor of five. Before performing an object to object comparison it is important to bear in mind that the category of `objects with IR excess' was assigned with different numerical criteria in the NIRI and NICMOS studies. While in the present study each object's IR excess was determined taking into account its uncertainty in the ({\it H-K$_s$}) color, in the NICMOS study a fixed limit of 0.4 mag was set as the minimum shift for all objects to flag them as showing an IR excess. This limit was chosen considering the maximum uncertainties in the photometry of the objects included in NICMOS sample.
Our comparison therefore focused on the 2 out of 12 objects selected as candidate MYSOs in the NICMOS study that are not among the objects with IR excess in Table \ref{tab:latablaIR}. The object located at $\alpha$=23.64050, $\delta$= +30.78196 lies extremely close to a RSG candidate which saturates in the NIRI images and might have hampered the chances of a clear detection in our photometry. The object located at $\alpha$=023.63578, $\delta$= +30.78483 is a faint source in a highly crowded area, two factors that affected the object detection in NIRI images. An inspection of the colors for the 10 objects with IR excess in common in both samples shows that the objects exhibit differences particularly in the ({\it J-H}) color, mainly caused by a combination of dissimilar transmission curves between the NIRI-{\it J} and NICMOS-F110W filters, and the fact that several of these objects were close to the detection limit in the F110W filter.

\begin{figure}
 \includegraphics[angle=-90, width=0.49\textwidth]{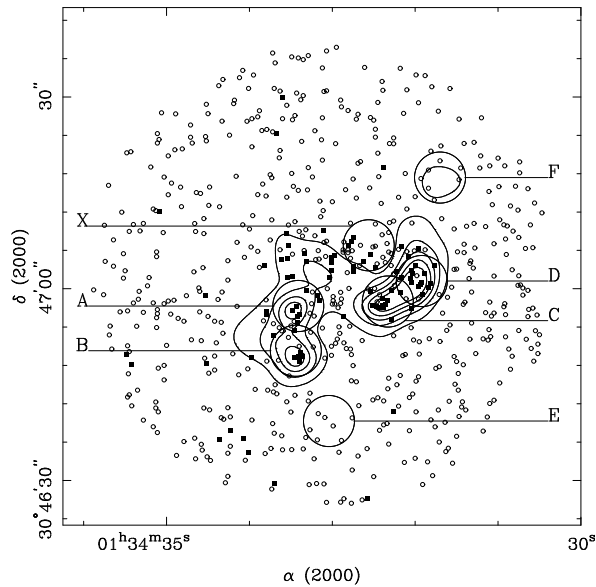}
 \caption{Spatial distribution of observed objects in the NGC\,604 central region. Radio continuum contours and regions as identified by \cite{1999ApJ...514..188C} are included. Open circles are stars without IR excess and filled squares denote the objects with IR excess. }
   \label{fig:churchwell_reg}
\end{figure}

Figure \ref{fig:churchwell_reg} shows the objects' distribution in the NGC\,604 central region, where open circles are stars without IR excess and filled squares denote the objects with IR excess. Contours correspond to the 8.44 GHz radio continuum emission by \cite{1999ApJ...514..188C}, designated following the authors' notation and pink circles denote subregions for which $\eta_{IR}$ was calculated as it is described below. In Figure \ref{fig:churchwell_reg} it is evident that most of the objects which
present IR excess are located in the regions delineated by the radio continuum
emission contours.  In \cite{1999ApJ...514..188C} the authors concluded
that the most luminous knots require the equivalent of five to eight O5\,III
stars to account for their
ionization. An inspection of Figure \ref{fig:churchwell_reg} yields a number ranging from 6 to 15 embedded objects for each peak. The spatial distribution of our candidate
MYSOs is also in agreement with a previous study by \cite{2007ApJ...664L..27T} 
who, by means of CO molecular emission observations at CO (J\,=\,1-0) and CO (J\,=\,3-2), inferred that
 the region of the arc coincident with the radio continuum emission C-D and
 its extension to the East (see
Figure 3 of the mentioned paper) have the appropriate high density and
temperature conditions to favor and support ongoing massive star formation
processes.
In addition, for the region coincident with the 
radio emission peak at knot A, \cite{2004AJ....128.1196M}, based on an exhaustive study of the different
gas phases at NGC\,604 with HST and ground-based data, suggested that in this location, there 
is a filled H{\sc ii} region (in contrast with H{\sc ii} surfaces) that may be
witnessing induced star formation. Similar conclusions related to the embedded
star formation in NGC\,604 were reached in a study by \cite{2009ApJ...699.1125R}.

Regarding the distribution of objects with IR excess in particular regions within NGC\,604, we have also calculated $\eta_{IR}$ for seven subregions in the NGC\,604 central region (marked as large circles in Figure \ref{fig:churchwell_reg}). The chosen subregions are coincident with the knots of continuum radio emission at 8.44 GHz \citep{1999ApJ...514..188C}, where most of the IR excess sources are concentrated. The regions were named A-F, following the authors nomenclature. An extra subregion, `X', located in NGC\,604's main cavity was added. Most of the stars within X belong to the main component of the SOBA, and it covers an area with a concentration of evolved stars. The diameter of the subregions was set at $\sim$\,33 pc (8$^{\prime\prime}$), based on the typical sizes of massive star formation regions
and OB associations in the Galaxy. For each subregion, $\eta_{IR}$  was calculated following the same procedure applied for the whole region. There are 7 objects in the control field within the area of each subregion ($\sim$\,845 pc$^ 2$) and none with IR excess. Hence, 7 objects were subtracted from the total number of objects in each subregion to account for background/foreground contamination. The extinction towards the radio components A-F, ranges from A$_V$\,=\,2.8 to 1.7 mag \citep{1999ApJ...514..188C} that is equivalent to A$_K$\,$\sim$\,0.28\,-\,0.17 mag, which will not affect considerably the obtained $\eta_{IR}$ values. The results are summarized in Table \ref{tab:conteos_regiones} where Columns 2 and 3 list the center coordinates of each subregion, Column 4 contains the $\eta_{IR}$ values in percentage and, in parenthesis, the ratio of the number of objects with IR excess to the total number of objects within the subregion after background/foreground contamination correction. In regions X and C, there were subtracted 2 and 1 objects with IR excess, respectively, since they were known WR stars. This analysis shows that the subregions with higher $\eta_{IR}$ are, in decreasing order, knots D, C and B, and A. There is a noticeable gap between these subregions and the X, F and E which contain little or no objects showing IR excess at all. 
Although we are aware that these percentages were calculated with relatively
few objects, and are therefore very sensitive to small changes in
the magnitude limits or subregion sizes, the results just described are well
established. Regions C, D, and B are located within the arc pointed-out by \cite{2007ApJ...664L..27T} to exhibit ongoing star-formation mentioned before, whereas \cite{2004AJ....128.1196M} proposed that region A may be witnessing induced star formation. On the other hand, region F, coincident with `Cluster B' from \cite{1996ApJ...456..174H} has an age similar to the main central cluster in NGC\,604, and E is coincident with the SNR found by \cite{1980A&AS...40...67D}. 
If $\eta_{IR}$ correlates with the age of the subregions, as is the case for studies in Galactic star-forming regions (see \citealt{2000AJ....120.1396H},\citealt{2001ApJ...553L.153H}), the results of this analysis, based on the NIR photometric characteristics of individual objects, is consistent with the idea that most of the objects in regions A, B, C, and D belong to a younger generation than the main SOBA population, which is supported by the mentioned previous studies based on global observations of the gaseous component.

\begin{table}
\centering
\caption{Infrared excesses derived for individual knots of star formation}\vspace{0.5cm}
\label{tab:conteos_regiones}

{\small
 \begin{tabular}{ccccccc}
\tableline
Region&  $\alpha$(J2000)   &	$\delta$(J2000) & $\eta_{IR}$ (excIR/all)  \\ 
      &	  (\textdegree) &   (\textdegree)  &	(\%)       \\
\tableline
A	& 23.639208 & 30.782583  & 69 (9/13)  \\
B 	& 23.639333 & 30.780639  & 75 (6/8)   \\
C 	& 23.634708 & 30.782778  & 75 (12/16) \\
D 	& 23.633000 & 30.783667  & 88 (15/17) \\
E 	& 23.637708 & 30.777583  & 0  (0/0)   \\
F 	& 23.632083 & 30.788167  & 0  (0/1)  \\
X	& 23.635685 & 30.785212	 & 15 (4/26) \\
\tableline
\end{tabular}
}
\tablecomments{\twocolumn{Each region is identified according to \cite{1999ApJ...514..188C} nomenclature, their center coordinates listed in Columns 2 and 3, their IR excess ratio shown in Column 4.}}
\end{table}

\subsection{Individual Objects}
\begin{figure}
 \includegraphics[angle=-90, width=0.49\textwidth]{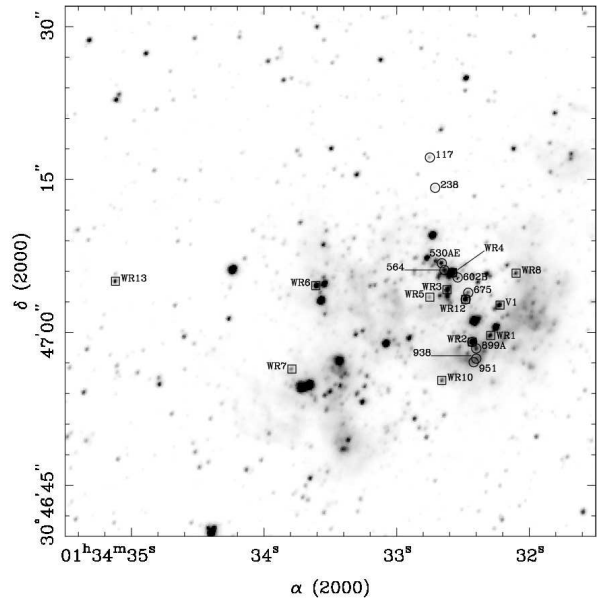}
 \caption{{\it J} band image with the identification of the stellar objects identified in the literature. \cite{2008MNRAS.389.1033D} sources are marked with square boxes and \cite{2003AJ....125.3082B} are shown with open circles. }
   \label{fig:ubicacion}
\end{figure}

\begin{figure*}
\includegraphics[width=0.5\textwidth]{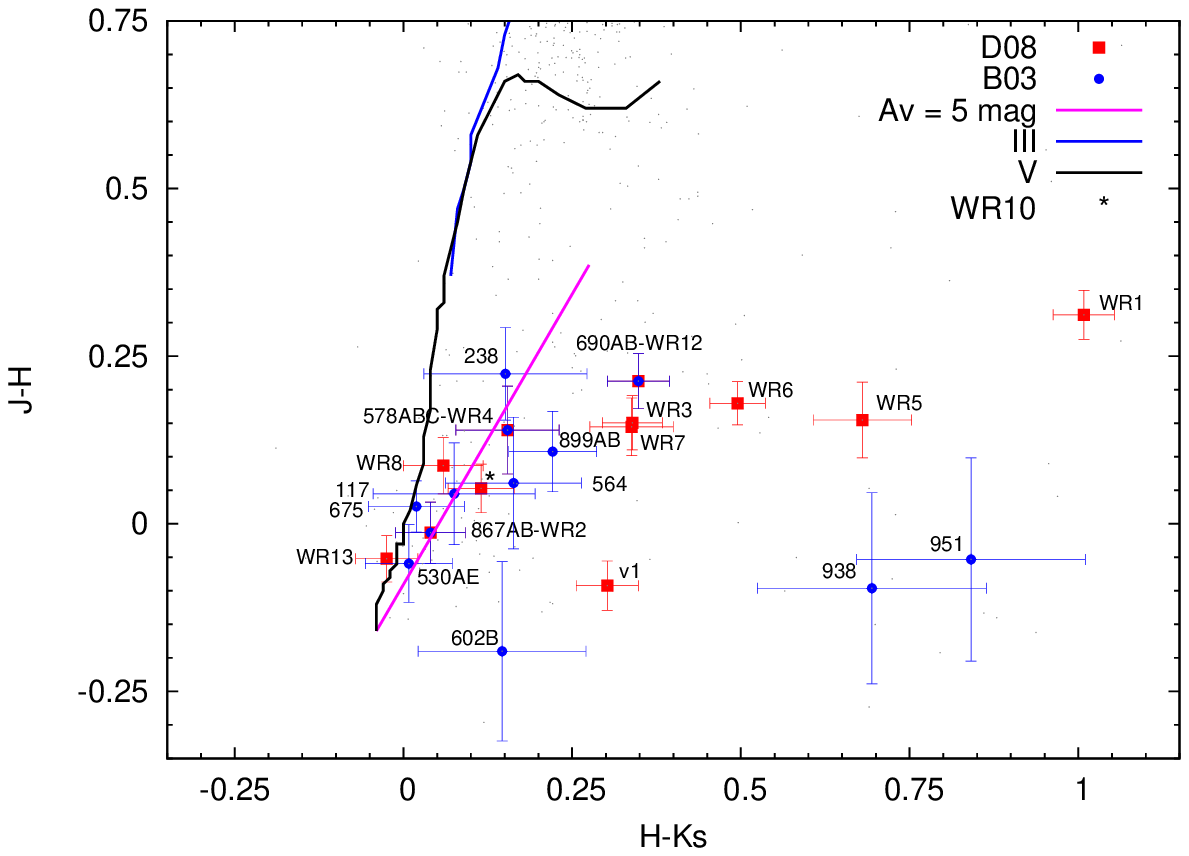}
\includegraphics[width=0.5\textwidth]{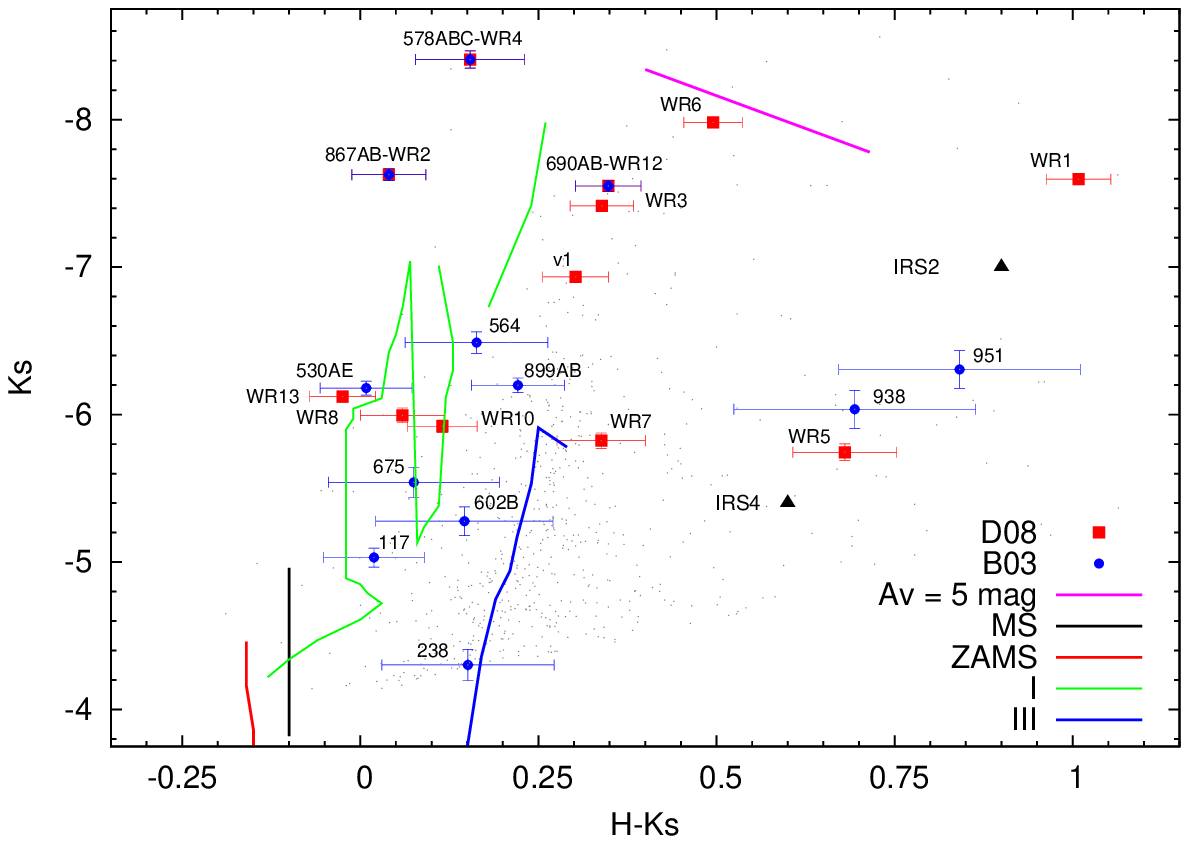}
 \caption{Color-color (left panel) and color-magnitude (right panel) diagrams for stars with known spectral types. Reference tracks, reddening lines and known massive young stellar objects are the same as those in Figures \ref{fig:CCD} and  \ref{fig:CMD}. Squares indicate objects listed by \cite{2008MNRAS.389.1033D} and circles those studied by \cite{2003AJ....125.3082B}. This Figure is presented in a color version in the electronic edition of the Astronomical Journal.}
   \label{fig:conocidas}
\end{figure*}

We have also made a brief survey of the relevant individual objects present in
our photometric study that were studied by other authors. 
These objects are identified in our NGC\,604 field image in Figure \ref{fig:ubicacion} 
and plotted in the respective CC and CM diagrams in Figure \ref{fig:conocidas}.
Circles show WR stars included in the study of \cite{2008MNRAS.389.1033D} and the references therein. Square boxes identify objects resulted from a visual identification between our NIRI images and the UV images from the study by \cite{2003AJ....125.3082B}. These objects are mainly OB stars with spectral classification derived by the authors from UV HST-STIS spectra.
The identification between the HST finding chart and our NIRI frames must be
considered with the caveat that, at M\,33's distance, for some objects we might be 
looking at unresolved stellar blends. An illustrative example can be found when comparing with analog regions, such as the star forming knots in 30 Doradus where \cite{1999AJ....117..225W} identified several strong knots lying 2-3 pc away from massive O3-6 stars. A similar configuration would fit within one arcsecond in our images.
 In Table
\ref{tab:known_comments} we have listed these objects, where Column 1 is the Id.
given in previous studies preceded by a set of characters that identify the
reference, Column 2 contains the object classification from the previous
studies, in Columns 3, 4, and 5 we have included our measured {\it J}, {\it H}, and {\it K$_s$} 
magnitudes, respectively. Those objects marked with $^c$ in Column 1 have a
special comment in the following paragraph.\\

\begin{table*}
\centering
\caption{Stars with known spectral types}\vspace{0.5cm}\label{tab:known_comments}
 \begin{tabular}{cllccc}
 \tableline
  Id &	Reference & Classification &	{\it J} & {\it H} & {\it K$_s$} \\
\tableline
3791 & D08-WR1         & WCE           & 18.32 $\pm$ 0.03 & 18.01 $\pm$ 0.03 & 17.00 $\pm$ 0.04 \\
1168 & D08-WR2 $^c$    & WN            & 17.00 $\pm$ 0.03 & 17.01 $\pm$ 0.03 & 16.97 $\pm$ 0.04 \\
1168 & B03-867 A-B$^c$ & O4Iab-O4Ia    &                  &	             &                  \\
3317 & D08-WR3         & WN            & 17.68 $\pm$ 0.03 & 17.52 $\pm$ 0.03 & 17.19 $\pm$ 0.04 \\
3795 & D08-WR4$^c$     & WN            & 16.49 $\pm$ 0.04 & 16.35 $\pm$ 0.05 & 16.19 $\pm$ 0.06 \\
3795 & B03-578A-C$^c$  & O9II-O9Ia-09II&                  &                  &                  \\
1257 & D08-WR5         & WC6           & 19.69 $\pm$ 0.03 & 19.54 $\pm$ 0.05 & 18.86 $\pm$ 0.06 \\
1265 & D08-WR6         & WNL           & 17.29 $\pm$ 0.02 & 17.11 $\pm$ 0.02 & 16.62 $\pm$ 0.03 \\
4021 & D08-WR7         & WC4           & 19.26 $\pm$ 0.03 & 19.12 $\pm$ 0.03 & 18.78 $\pm$ 0.05 \\
4096 & D08-WR8$^c$     & WN6           & 18.75 $\pm$ 0.03 & 18.66 $\pm$ 0.03 & 18.61 $\pm$ 0.05 \\ 
3255 & D08-WR10        & WN6           & 18.85 $\pm$ 0.03 & 18.79 $\pm$ 0.03 & 18.68 $\pm$ 0.04 \\
4075 & D08-WR12$^c$    & WNE           & 17.61 $\pm$ 0.03 & 17.40 $\pm$ 0.03 & 17.05 $\pm$ 0.04 \\
4075 & B03-690A-B$^c$  & O5III-BOIb    &                  &                  &                  \\    
1247 & D08-WR13$^c$    & O6.5Iaf       & 18.40 $\pm$ 0.02 & 18.45 $\pm$ 0.03 & 18.48 $\pm$ 0.04 \\
1244 & D08-V1          & Of/WNL        & 17.88 $\pm$ 0.02 & 17.97 $\pm$ 0.03 & 17.67 $\pm$ 0.04 \\
1631 & B03-117         & O4II          & 19.61 $\pm$ 0.02 & 19.59 $\pm$ 0.03 & 19.57 $\pm$ 0.06 \\
1545 & B03-238         & O9III:        & 20.67 $\pm$ 0.04 & 20.45 $\pm$ 0.06 & 20.30 $\pm$ 0.11 \\
1327 & B03-530A-E      & O7V:          & 18.37 $\pm$ 0.04 & 18.43 $\pm$ 0.04 & 18.42 $\pm$ 0.05 \\
4297 & B03-564$^c$     & O9II          & 18.34 $\pm$ 0.07 & 18.28 $\pm$ 0.07 & 18.11 $\pm$ 0.07 \\
4380 & B03-602B        & O9II          & 19.28 $\pm$ 0.11 & 19.47 $\pm$ 0.08 & 19.32 $\pm$ 0.10 \\
4362 & B03-675         & O7II          & 19.18 $\pm$ 0.04 & 19.13 $\pm$ 0.06 & 19.06 $\pm$ 0.10 \\
4224 & B03-899A        & O7II          & 18.73 $\pm$ 0.04 & 18.62 $\pm$ 0.04 & 18.40 $\pm$ 0.05 \\
4425 & B03-938$^c$     & O6III::       & 19.16 $\pm$ 0.09 & 19.26 $\pm$ 0.11 & 18.56 $\pm$ 0.13 \\
4335 & B03-951$^c$     & O8V           & 19.08 $\pm$ 0.10 & 19.14 $\pm$ 0.11 & 18.29 $\pm$ 0.13 \\
\tableline
 \end{tabular}
\tablecomments{Column 1 shows the object id.\ from this work, Column 2 includes different identifiers used in the literature for the same object, Column 3 lists the spectral classification and Columns 4 to 6 include NIR magnitudes derived in the present work. Those objects marked with $^c$ in Column 1, have a special comment in this Section. D08: \cite{2008MNRAS.389.1033D}, B03: \cite{2003AJ....125.3082B}, the
authors have identified with same number and a different character those object
with overlapping spectra.}
 \end{table*}

\noindent{\bfseries Comments on individual objects:}\\
WR stars tend to be red objects in the NIR CC and CM diagrams since they are
more luminous in the {\it K} band than the {\it J} band as they are being observed
through their own obscured envelope and winds. 

D08-WR2 was first classified as WN by \cite{1981ApJ...249..471C}, who could not 
resolve it from D08-WR1. D08 did not include this
object in their sample. B03 classified it as O4Iab-O4Ia, since they could
identify two objects at this position, being the most massive and luminous as
well as the youngest objects of their sample. From a visual inspection of our
images, it is evident that there is more than one object at this location
although we are not able to resolve them photometrically. The  ({\it H-K$_s$})
color of this object in our photometry is bluer than the expected for a WN
object but also its {\it K$_s$} magnitude is brighter than the magnitude of a single O
I star, which can be explained 
as the composite
emission from two (or more) unresolved objects, in agreement with B03
observations.

D08-WR4 presents a similar situation as D08-WR2. This object was
classified as WN by \cite{1981ApJ...249..471C}, who could not 
resolve it from D08-WR3, and D08-WR5. D08 did not obtain a new spectrum for it, but at
the same location B03 obtained a spectra and classified three objects: two are O9II stars and a O9Ia. Again, in our images we are able to
resolve only one object but a visual inspection suggests that at this location
there is more than a single object. Besides, the blue shifted  ({\it H-K$_s$}) color
and the bright {\it K$_s$} magnitude indicate that we are possibly measuring the
composite emission of a tight group of massive stars (not necessarily evolved as
WR objects) in agreement with B03 classification.

D08-WR6 is one of the brightest objects in the region. It was first observed by \cite{1993AJ....105.1400D} and studied later by \cite{1996MNRAS.279.1219T}, who measured a He overabundance and extremely wide spectral emission lines ($\sim$\,2500 km s$^{-1}$) as well as a spectral variability in a time span of $\sim$\,10 years. The authors classified this object as a transition object between a Luminous Blue Variable and a WR star.

D08-WR8 was already known as a WR object candidate when D08 precisely
classified it as WN6. In the HST F170W images it is clear that there are two
objects very close together in projection, although our photometry distinguishes only one
object, that may cause the D08-WR8 blue color in our CC diagram. 

D08-WR12 is resolved into two objects in B03 HST images, the authors
classified them as O5III and a B0Ib, although D08 is in disagreement with these
spectral types, based mainly on the presence of the He{\sc ii} ($\lambda$4686) emission
line. D08 argued that although their spectrum is a composite of two stars most
of the emission lines must come from the brightest star (the B0Ib for B03) and
they classified it as a WN10 star.

D08-WR13, this object is approximately 30 arcseconds away from the central cluster. D08 suggested that its spectral type is O6 Iaf. This object does not exhibit IR excess in our photometry.

B03-564 is really two stars but we are not able to resolve them, so our magnitudes
have contributions from both objects.  

B03-938 was classified as O6III:: by B03. In the HTS-STIS F170W image it
is clear that there are two objects close together in projection while in our
photometry only one is detected. Besides, these objects are located in a region
within a ring of noticeable nebular emission which probably causes the red color of this object in our CC diagram.

B03-951, is located close to B03-938, in the same nebular ring, and it is also red for its O8\,V spectral type according to B03 classification. 

\section{SUMMARY AND CONCLUDING REMARKS}
\label{conclusions}
We have performed a NIR photometric study on the NGC\,604 massive star-forming region, using Gemini-NIRI images.

Our study was focused on the detection and individual identification of massive young stellar objects that comprise a new stellar generation in the region.

By means of {\it JHK$_s$} photometry we have identified 68 candidates MYSOs. These objects were selected by their position in the CC diagram (each of these objects exhibits IR excess greater than their photometric error in  ({\it H-K$_s$})).

The analysis of the spatial distribution shows that most of the MYSOs candidates are located in areas with strong nebular emission, delineated by the radio continuum contours at 8.44 GHz, where previous studies stated the possibility of star formation processes taking place.

We have calculated the fraction of objects with IR excess throughout the whole region. The result yields that several regions in NGC\,604 show a high fraction of objects with IR excess for the cluster age. Although it is not possible to perform a direct comparison with star forming regions in our Galaxy, this could be interpreted as a recent burst of star formation. Studies of other GHRs are necessary to establish whether there is a general
tendency for regions of massive star formation to present higher IR-excess
fractions and to understand the true nature of IR-excesses in MYSOs.

We have made a short review of individual objects in NGC\,604 included in our sample, which were already studied by other authors.

All these results complete a general picture that exhibits the existence of a new stellar generation in the NGC\,604 region, which is mainly taking place in the molecular gas that extends to the south-east side of the main SOBA population.

This is the first study of the NGC\,604 star forming region dedicated to the detection of individual MYSOs candidates. New detailed observations for each object are needed to confirm the nature of the MYSOs candidates we have found and also to provide information on the physical conditions of the circumstellar matter which is causing the observed IR excess.\\

Acknowledgments

This research was based on observations obtained at the Gemini Observatory, which is operated by the 
Association of Universities for Research in Astronomy, Inc., under a cooperative agreement 
with the NSF on behalf of the Gemini partnership: the National Science Foundation (United 
States), the Science and Technology Facilities Council (United Kingdom), the 
National Research Council (Canada), CONICYT (Chile), the Australian Research Council (Australia), 
Minist\'{e}rio da Ci\^{e}ncia e Tecnologia (Brazil) 
and Ministerio de Ciencia, Tecnolog\'{i}a e Innovaci\'{o}n Productiva (Argentina). The authors would like to specially thank Andrew Stephens, from Gemini Observatory, for his help regarding the data processing procedures of Gemini-NIRI images and Andrew McWilliam for his final revision of the English text.
 We have also benefited from fruitful discussions with Roberto Terlevich. We are grateful to the anonymous referee for the thorough review which has led to a much improved version from the original manuscript.

\bibliographystyle{apj}
\bibliography{biblio_ngc604}

\label{lastpage}
\end{document}